\documentclass[conference]{IEEEtran}
\IEEEoverridecommandlockouts
\usepackage[noadjust]{cite}

\usepackage{graphicx}
\usepackage{amsmath,amssymb,amsfonts}
\usepackage[capitalise]{cleveref}

\usepackage{algorithm}
\newcommand{\algparbox}[1]{\parbox[t]{\dimexpr\linewidth-\algorithmicindent}{#1\strut}}
\usepackage{algpseudocode}
\algnewcommand\algorithmicinput{\textbf{Input:}}
\algnewcommand\algorithmicoutput{\textbf{Output:}}
\algnewcommand\Input{\item[\algorithmicinput]}%
\algnewcommand\Output{\item[\algorithmicoutput]}%
\let\oldnl\nl
\newcommand{\nonl}{\renewcommand{\nl}{\let\nl\oldnl}}
\usepackage{bm}
\usepackage{enumerate}
\usepackage{enumitem}
\usepackage{textcomp}
\usepackage{xcolor}
\usepackage[normalem]{ulem}
\def\BibTeX{{\rm B\kern-.05em{\sc i\kern-.025em b}\kern-.08em
    T\kern-.1667em\lower.7ex\hbox{E}\kern-.125emX}}
\usepackage[papersize={8.5in,11in}, left=0.625in, right=0.625in, top=0.75in, bottom=1in]{geometry}
\usepackage{acronym}
\acrodef{COP}{Configuration and Optimization Parameter}
\acrodef{QoE}{Quality of Experience}
\acrodef{CIO}{Cell Individual Offset}
\acrodef{HOM}{Handover Margin}
\acrodef{TTT}{Time-to-Trigger}
\acrodef{SINR}{Signal to Interference and Noise Ratio}
\acrodef{KPI}{Key Performance Indicator}
\acrodef{GA}{Genetic Algorithm}
\acrodef{ML}{Machine Learning}
\acrodef{COPs}{Configuration and Optimization Parameters}
\acrodef{KPIs}{Key Performance Indicators}

\begin{document}

\title{A Machine Learning based Framework for KPI Maximization in Emerging Networks using Mobility Parameters}
% {\footnotesize \textsuperscript}
% }

\author{Joel Shodamola, Usama Masood, Marvin Manalastas and Ali Imran\\
AI4Networks Research Center, Dept.\ of Electrical \& Computer Engineering, University of Oklahoma, USA\\
Email: \{joelshodams, usama.masood, marvin, ali.imran\}@ou.edu

}

\maketitle

\begin{abstract}Current LTE network is faced with a plethora of \ac{COPs}, both hard and soft, that are adjusted manually to manage the network and provide better \ac{QoE}. With 5G in view, the number of these \ac{COPs} are expected to reach 2000 per site, making their manual tuning for finding the optimal combination of these parameters, an impossible fleet. Alongside these thousands of \ac{COPs} is the anticipated network densification in emerging networks which exacerbates the burden of the network operators in managing and optimizing the network. Hence, we propose a machine learning-based framework combined with a heuristic technique to discover the optimal combination of two pertinent \ac{COPs} used in mobility, \ac{CIO} and \ac{HOM}, that maximizes a specific \ac{KPI} such as mean \ac{SINR} of all the connected users. The first part of the framework leverages the power of machine learning to predict the \ac{KPI} of interest given several different combinations of \ac{CIO} and \ac{HOM}. The resulting predictions are then fed into \ac{GA} which searches for the best combination of the two mentioned parameters that yield the maximum mean \ac{SINR} for all users. Performance of the framework is also evaluated using several machine learning techniques, with CatBoost algorithm yielding the best prediction performance. Meanwhile, \ac{GA} is able to reveal the optimal parameter setting combination more efficiently and with three orders of magnitude faster convergence time in comparison to brute force approach.

\end{abstract}

\vspace{-10pt}
\section{Introduction}
Since its advent, the wireless network has undergone several phases of technological advancement. From 2G technology which mainly supports voice communication to 4G which brought us high speed internet and now to upcoming 5G which promises ultra-fast connection speed and massive capacity. However, alongside this development is the increase in intricacy of the cellular network. As a rule of thumb, the more services a network supports, the more complex the functionalities and configuration parameters of the network become. This rise in complexity can be mirrored by the fact that the number of \ac{COPs} per site has risen steadily from 500/site for 2G to 2000/site for 5G.

These \ac{COPs} directly impact the networks performance and are usually measured through \ac{KPIs}. \ac{COPs} that are sub-optimally tuned usually lead to poor network performance and degraded KPIs which ultimately result to unsatisfactory user \ac{QoE}. That is why it is critical to make sure that these parameters are correctly set and adjusted. Current industry practice is to tune these parameters manually based on domain knowledge, intuition and sometimes hit and trial approach. However, manual configuration of these numerous parameters can be time consuming, inefficient and expensive on the side of the operators. As a result, mobile operators tend to deal on tweaking a limited number of \ac{COPs} to improve certain \ac{KPIs}, therefore limiting the possibility of deriving the utmost performance that a network can achieve. In addition, the expected densification in emerging network will add to this problem which will make optimal parameter setting discovery unfathomable.

In order to tune these parameters properly, it is essential to learn how KPIs behave with the variations in \ac{COPs}. However, as operators try very little combinations of configuration parameters in real networks, it is very difficult to interpolate and learn the behavior of the system performance due to changes in COPs. For this reason, the task of modeling COP-KPI relation is almost impossible using conventional interpolation techniques. This is where \ac{ML} comes into play. With machine learning, it is possible to model and map out functions that cannot be directly or mathematically interpreted in the data \cite{klaine2017survey,masood2019machine}. This capability makes machine learning a promising tool to accurately capture the network dynamics due to changes in COPs even with very little and sparse experiment data.

To overcome the current challenges of learning COP-KPI relations and finding optimal \ac{COP} combination to maximize \ac{KPIs}, we have proposed a framework that involves: first, leveraging the power of machine learning to learn and predict the network performance given a set of \ac{COP} using data gathered from a realistic industry-grade simulator and second, finding the optimal \ac{COP} combination values that will yield the maximum \ac{KPI} using a heuristic search technique in the form of \ac{GA}. In this paper, two pertinent \ac{COPs} related to mobility, \ac{CIO} and \ac{HOM} are chosen, with mean \ac{SINR} as the optimized \ac{KPI}.

\subsection{Relevant Work}

3GPP \cite{3gpp2011evolved} introduced the \ac{CIO} parameter as an LTE standard that controls handover and ensures proper load balancing from one cell to another. Since then, \ac{CIO} has been a parameter of interest to the research community to find out the optimal values of \ac{CIO} that lead to a desired \ac{KPI}. However, sole optimization of \ac{CIO} also generated further challenges in cell management resulting to a drop in \ac{SINR} \cite{asghar2018concurrent}. It is therefore pertinent that \ac{CIO} is combined with other mobility parameters like hysteresis and \ac{TTT} to taper off mobility related problems such as handover failure cases like too early, too late and handover to wrong cells.

Authors in \cite{munoz2013potential} observe the effect of \ac{HOM} and \ac{TTT} on call drop ratio (CDR) and handover ratio and conclude that the effect of \ac{TTT} is inconsequential. However, the underlying task is not only to discover what particular set of parameters has more influence on a desired \ac{KPI} but the set of values of these parameters that produce the optimal KPI. Authors in \cite{ewe2011base} determine optimal values of \ac{TTT} and \ac{CIO} from initial suboptimal configurations based on weighting factors with respect to handover failure cases. In \cite{kitagawa2011handover}, authors discuss the optimal values of \ac{HOM} with respect to user mobility and handover failure rates while authors in \cite{hegazy2015user} use \ac{TTT} and \ac{HOM} to address the same issue. However, none of these aforementioned studies leveraged on machine learning to derive optimal values of parameters.

Most of the relevant studies either used machine learning (ML) techniques to tune hard parameters or used a single \ac{ML} algorithm for \ac{KPI} optimization. Authors in \cite{lin2019machine} used \ac{ML} to optimize the same \ac{KPI} in our study but with reference to hard parameters like antenna azimuth. Although authors in  \cite{roshdy2018mobility} used machine learning, support vector machine (SVM) in particular, they tuned one handover parameter, \ac{CIO} to enhance user throughput. In \cite{ahmed2018optimization}, authors use \ac{GA} without any application from \ac{ML} to determine optimal values of just \ac{HOM} to attain reduction in power consumption.

In this paper, \ac{ML} is utilized to capture the \ac{KPI} behavior with changes in COPs before applying \ac{GA} to maximize the objective function. In that light, the main contributions of this paper are:
\begin{itemize}
       \item A novel combination of mobility parameters consisting of \ac{CIO} and \ac{HOM} as \ac{COPs} for \ac{COP}-\ac{KPI} modeling using machine learning
       \vspace{5pt}
       \item Performance comparison of different \ac{ML} algorithms for \ac{COP}-\ac{KPI} modeling
       \vspace{5pt}
       \item Use of genetic algorithm on the outputs of \ac{ML} model for faster convergence to achieve optimal \ac{COP} combination
\end{itemize}
      
\subsection{Parameters Definition}

\begin{figure}[!h]
\centerline{\hspace*{1cm}\includegraphics[scale=0.3]{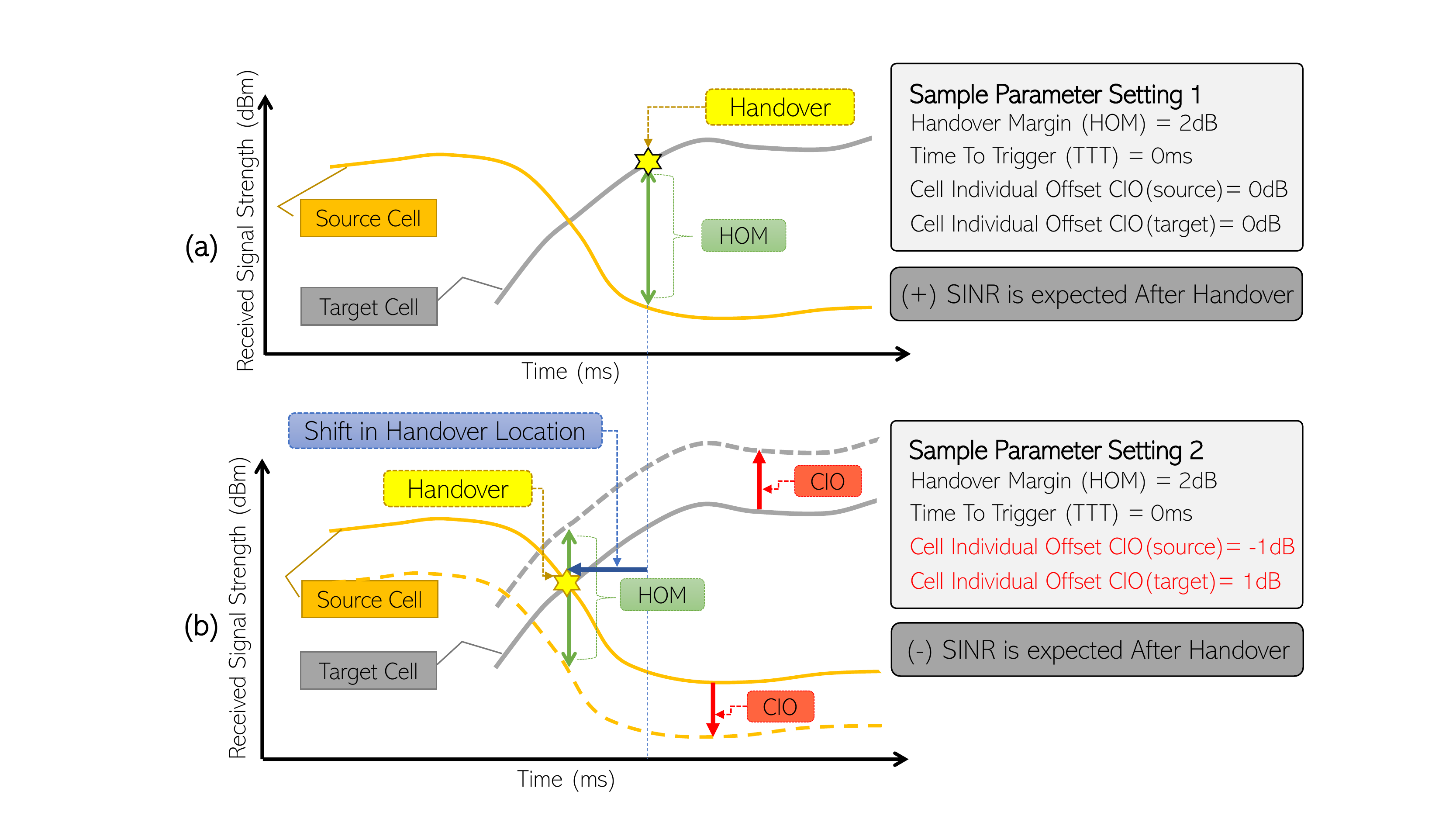}}
\caption{Effect of \ac{CIO} on Handover and \ac{SINR}}
\vspace{-15pt}
\label{fig1}
\end{figure}

\begin{enumerate}
\item \uline{Cell Individual Offset (CIO):} This is a measurement quantity used to determine cell association, regulate cell coverage and usually incorporated in power measurement to control handover. As shown in \cref{fig1}, handover can be made earlier (or later) by changing the \ac{CIO} values. \ac{CIO} is usually given positive or negative values; where a positive value augments the RSRP of the cell and a negative value diminishes the RSRP of intended cells. Fig. 1 also shows the effect of suboptimal \ac{CIO} tuning in terms of SINR. As the figure shows, wrongly tuned \ac{CIO} can cause too early handover where the source cell is still better than the target cell resulting to high interference and thus poor SINR.\
\vspace{2pt}
\item \uline{Handover Margin (HOM):} \ac{HOM} is a mobility related parameter that administers handover. As shown in Fig. 1(a) and(b), it is the assigned threshold that must be met, when the power of a target cell begins to exceed the power of the serving cell. It is usually within the range of 0 to 10dB.\
 \vspace{2pt}
\item \uline{Capacity:} We can define cell capacity in terms of SINR. \ac{SINR} is mostly used for radio link quality and throughput measurements. With throughput being the maximum data rate within a channel and from shannon capacity in equation (2), increase in \ac{SINR} will lead to a proportional increase in capacity. \cite{asghar2018concurrent} described this relationship in expression (2). The mean \ac{SINR} of a user associated with a cell \textit{s} for a given time slot t is:
\begin{equation}
\gamma_{x}^{s}=\frac{P_{t}^{s}d_{x}^s}{\Sigma_{1}^{i}w_{i}P_{t}^{i}d_{x}^{i}+N_p}\
\end{equation}

\begin{equation}
Capacity = N_{s} - \frac{1}{w_{i}}\sum_X\frac{\tau_x}{log_{2}(1+\gamma_{x}^{s})}
\end{equation}

where $\gamma_{x}^{s}$ is the average \ac{SINR} of user \textit{x} and cell \textit{s}, with $P_{t}^{s}$ and $P_{t}^{i}$ as transmission powers of cell \textit{s} and interfering cell \textit{i}, $d_{x}^{s}$ and  $d_{x}^{i}$ as the pathloss components of associated cell \textit{x} and interfering cells \textit{i} respectively. N is the noise, $w_{i}$ the load of the interfering cell and $N_{s}$ the total physical resource blocks (PRB).
\end{enumerate}

The rest of this paper is organized as follows: Section II highlights the proposed framework, Section III discusses the simulation setup. In Section IV, results and analysis are presented  and finally Section V concludes the paper.

\begin{figure*}[!t]
\vspace{-10pt}
\centerline{\includegraphics[trim=0 300 0 0,clip=true,scale=0.6]{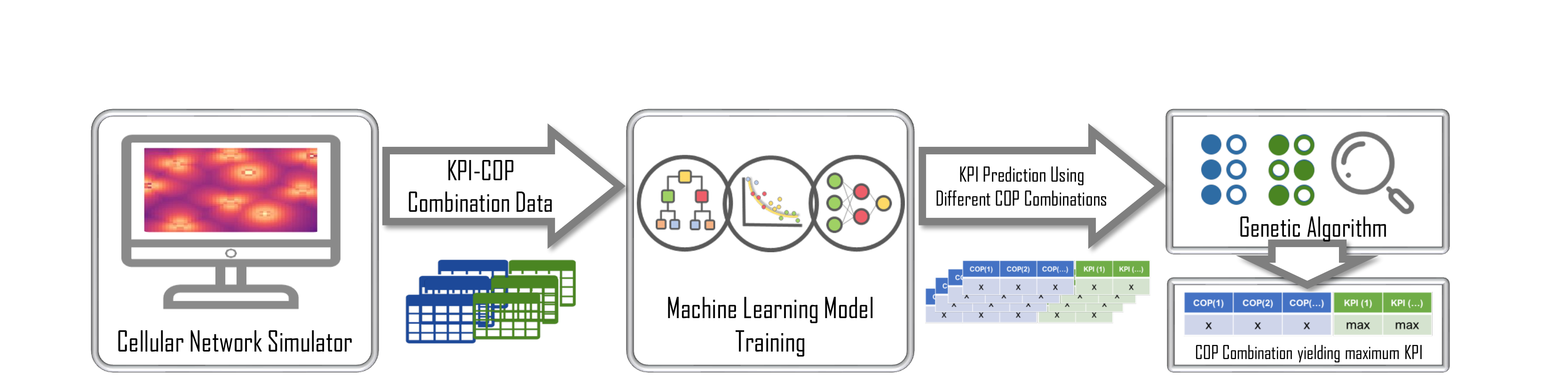}}

\caption{Proposed Framework for \ac{KPI} Maximization}
\label{fig2}
\end{figure*}
\section{System Framework}

As shown in Fig. 2, the proposed system framework takes into consideration three processes to achieve the objective function. First is the data generation part. For this, an industry grade system level simulator is used. Simulators have the advantage of generating large amount of data such as COP-KPI relations which are otherwise impossible to gather from a real network. The second part of the framework is the \ac{KPI} prediction using machine learning techniques. Here, \ac{COP} combination and \ac{KPI} data gathered from the simulator are used to train the machine learning models and learn how changes in \ac{COP} combinations affect the KPIs. This prediction model enables network operators  to envisage network behaviours and extrapolate important information. The last part of the framework involves discovery of the optimal parameter values using genetic algorithm. In this stage, prediction model of best performing machine learning model is used to initialize parameters. Detailed discussion of the data generation, machine learning techniques used, and how genetic algorithm is leveraged is shown below.

\subsection{Data Generation}
In order to learn how the cellular network behaves with changes in \ac{COP} values, one approach is to try several combinations of these parameters and then observe the resulting \ac{KPI} trends. However, this approach is not practical especially in live networks as there are hundreds of thousands of parameter combinations possible. Aside from that, one badly tuned parameter might impose a huge risk in the network’s performance. For this reason, simulators are a viable alternative to gather data which cannot be done on a live network due to the above-mentioned restrictions.

We wrote a script to automate the data generation process. This script automatically changes the values of COPs and records the resulting \ac{KPI} for each combination. The output of the simulator coupled with the automation script is a table of different \ac{CIO} and \ac{HOM} combinations with the expected \ac{SINR} for each combination. The data is utilized to train several machine learning models and also to predict the average \ac{SINR} for each CIO-HOM combination. Models are trained in two types of training data, one using all possible combinations (100 percent) and one using only (10 percent) of the training data. This is done to test the robustness of the machine learning model to sparse data which is usually the case in a real cellular network.  

\subsection{Machine Learning Algorithms for \ac{KPI} Prediction}

Machine Learning can be categorized as supervised and unsupervised learning. The former requires a set of training data  to learn from and maps to a corresponding output based on observed relationship from the trained data while the latter takes unlabelled data and understands its distribution to map into several labels. Supervised learning can be further categorised as classification and regression problems. The linearity or non-linearity of these relationship are usually represented and this differs from model to model. We apply various \ac{ML} algorithms and observe  their  prediction performance. In this paper, five machine learning models are evaluated.

\subsubsection{Linear Regression}
% {i. \textit{Linear Regression: }}
This method is an example of a solution to a regression based problem which describes the functional relationship between dependent ($y_{i}$) and independent attributes $(x_{i}$) by ascribing weights, \textit{w} to the function through a method of linearity as shown in equation (4). The target attribute (dependent variable), is extracted from a set of N input features  with coordinates ${(x{_i}, y{_i})}^N_{i=1}$ and $f(x{_i})$ being a function  of the weights $\phi(x{_i}, w)$\cite{Andriy2019Burkov}  . This can be summarized in equation (3) where $\epsilon$ is the error. 
\begin{equation}
\vspace{-0.3cm}
y_{i}=f(x_{i}) + \epsilon  \label{eq1}
\vspace{-0.3cm}
\end{equation}

\begin{equation}
y_{w,b} = wx +b  \label{eq}
\end{equation}
 
\subsubsection{K-Nearest Neighbor (KNN)}
% {ii. \textit{K-Nearest Neighbor (KNN)}:}
KNN makes no assumption for distribution of vectors by using distance calculation measures to categorize a set of input features in vector space based on a variable, k, which symbolizes the number of neighbors or surrounding features. Essentially, as batches of new independent features are introduced,  it locates the k neighbors closest to the new features and ascribes them to a target majority. The major distance metrics are cosine similarity and euclidean distance.
\subsubsection{Extreme Gradient Boosting (XGBoost)}
% {iii. \textit{Extreme Gradient Boosting (XGBoost):}}
This is an instance of a category of  gradient boosting called ensemble learning  which  aggregates multiple weak models (trees), iterates through them by ascribing weights to each model to give a final model. Rather than some random selection, each subsequent model is a correction of the previous. Two objective parameters are introduced here namely training loss and regularization function as shown in equation (5). 
\vspace{-0.1cm}
\begin{equation}
obj(\theta)=L(\theta) + \omega(\theta). \label{eq}
\vspace{-0.1cm}
\end{equation}
Here \textit{L} represents training loss and $\omega$, regularization. XGBoost adds some extra functions to reduce computational complexity and improve on scalability and prediction accuracy. 

\subsubsection{Categorical Boosting (CatBoost)}
% {iv. \textit{Categorical Boosting (CatBoost):}}
This is another category of gradient boosting which falls under ensemble methods. CatBoost was developed to give a better performance than its contemporaries like XGBoost and LightGBM in terms of efficiency, reducing the time wasted on tuning of parameters through excellent default functions and reduces overfitting on training samples. It uses an ordered mechanism to get training samples per time.

\subsubsection{Deep Neural Network (DNN)}
% {v. \textit{Deep Neural Network (DNN)}}:
DNN is a class of Artificial Neural Network (ANN) inspired by the biological brain that consists of hidden layers between input and output, represented by multi-perceptrons to discover hidden features and layers of inputs. Each hidden layer is supported by activation functions among which are TanH, Sigmoid and ReLu  depending on the nature of the problem. The outputs also are supported by optimizers namely Stochastic Gradient Descent (SGD), AdaDelta, AdaGrad and Adam with the latter being the most efficient. DNN is able to learn intricate characteristics of input vectors and categorize them more efficiently.\

\subsection{Genetic Algorithm}\label{AA}
\begin{figure}[htbp]
\vspace{-10pt}
\centering{\includegraphics[scale=0.5]{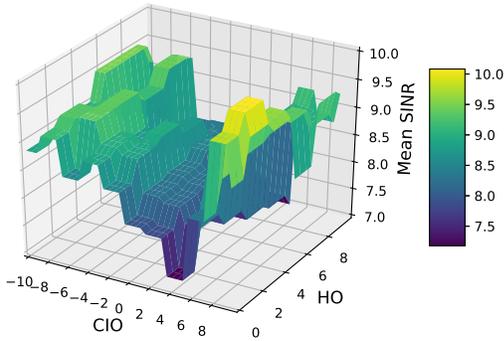}}
\caption{Non-convexity of Objective Function.}
\label{fig3}
\end{figure}

To show the intrinsic nature of the  relationship between objective function and parameters, a plot of \ac{SINR} as a function of \ac{CIO} and \ac{HOM} for one base station (BS) is shown in \cref{fig3}. From this figure, the problem is seen as a non-convex function which can also be termed as a NP-hard optimization problem with several local maxima and a global maxima. We use genetic algorithm (GA) in conjunction with machine learning model as the heuristic optimization technique based on its induction from natural and biological evolution.  

\begin{algorithm}
        \caption{Genetic Algorithm for Objective Function}\label{alg:rss}

        \begin{algorithmic}[1]
                \Input 
            \Statex Parameters X($CIO_{1}$, $CIO_{2}$, $CIO_{3}$, $HOM_{1}$, $HOM_{2}$, $HOM_{3}$)
                 \Statex Machine learning model  (\textbf{F})
                \Statex sample/iteration \textbf{P}
                \Statex Maximum Generation \textbf{G}
                \Statex Population \textbf{Pop} \Statex Crossover \textbf{Cr} \Statex Mutation \textbf{M}

                \Output 
                \Statex Solution $X_{sol}$ = X($CIO_{1}$, $CIO_{2}$, $CIO_{3}$, $HOM_{1}$, $HOM_{2}$, $HOM_{3}$)
                \Statex argMax[X, f(X)]\\
               Randomly generate parameters from X with size \textbf{P};\\
               Assign fitness function from \textbf{F} for each sample \textbf{P};\\
               open an empty set \textbf{POP} and save \textbf{P} inside;
            \For{i=1 to \textbf{G}}
            \State {evaluate \textbf{POP} for each \textbf{P} sample using \textbf{F};}
            \State {create an empty set called Parent \textbf{Par};}
            \State {select best elites from \textbf{POP} and save in \textbf{Par};} 
                    \For {two best elites in \textbf{Par}}
                    \State\algparbox{perform crossover \textbf{Cr} (Par1, Par2);}
                    \State \algparbox{get two children from \textbf{Cr} and mutate each;\\
                    \textbf{M}({Cr1,Cr2});}
                    \EndFor
            \State {Store each child solution and update in \textbf{Pop}};
			\EndFor\\
                \textbf{Return:} The last set of \textbf{POP} and that becomes $X_{sol}$
        \end{algorithmic}

\end{algorithm}
Essentially, \ac{GA} generates a population of a specified size that consists of several individuals (or solutions), each consists of variables that are randomly initialized. In our framework, these initial values are pulled from the \ac{ML} prediction function. A criteria is set for maximum iteration, where for each iteration, solutions are generated and passed to the subsequent generation. The algorithm keeps iterating until it converges to an optimal solution.  Each of the solutions in these generations are evaluated by a fitness function derived from the machine learning model to select elites considered as the best solutions or parents that are further iterated for crossover and mutation. Algorithm 1 shows the pseudo code with the detailed break down on how this \ac{GA} process occurs. The crossover method used was Simulated Binary crossover (SBX) which  converts each  values to binary values to initiate the single point crossovers. SBX incorporates a spread factor and makes a random selection from a probability distribution.

\section{Simulation setup}
Using the ray-tracing based industry grade system-level simulator, we have created a cellular network scenario composed of 12 macro base stations with 3 sectors each. However, changes in parameters are concentrated only to 3 sectors as shown in Fig. 4. The rest of the base stations are used to add interference and thus capture realistic SINR. A total of 356 users are created which are randomly distributed around the simulation area. To make it more realistic, these users have different speeds as well as different services used such as VoIP and high-speed internet. Table 1 shows the detailed parameter settings for the simulation. Each base station is equipped with two mobility parameters, \ac{CIO} and HOM. For the base stations of interest, the value of \ac{CIO} and \ac{HOM} ranges from [-10 to 10] dB and [0 to 10] dB respectively. For the rest of the base stations, the value of these parameters remains constant. With \ac{CIO} having 21 possible values, 11 for \ac{HOM} and considering three BSs, we would have $21^{3}\times11^{3}$ (12,326,391) possible \ac{COP} combinations. However, we have observed that using all this possible combination is not necessary as \ac{SINR} values change minutely. Therefore, we have decided to use \ac{CIO} and \ac{HOM} values with step size of 2 to reduce the data size, that is ($11^{3}\times6^{3}$) possible combinations.

\begin{figure}[htbp]
\centerline{\includegraphics[scale=0.5]{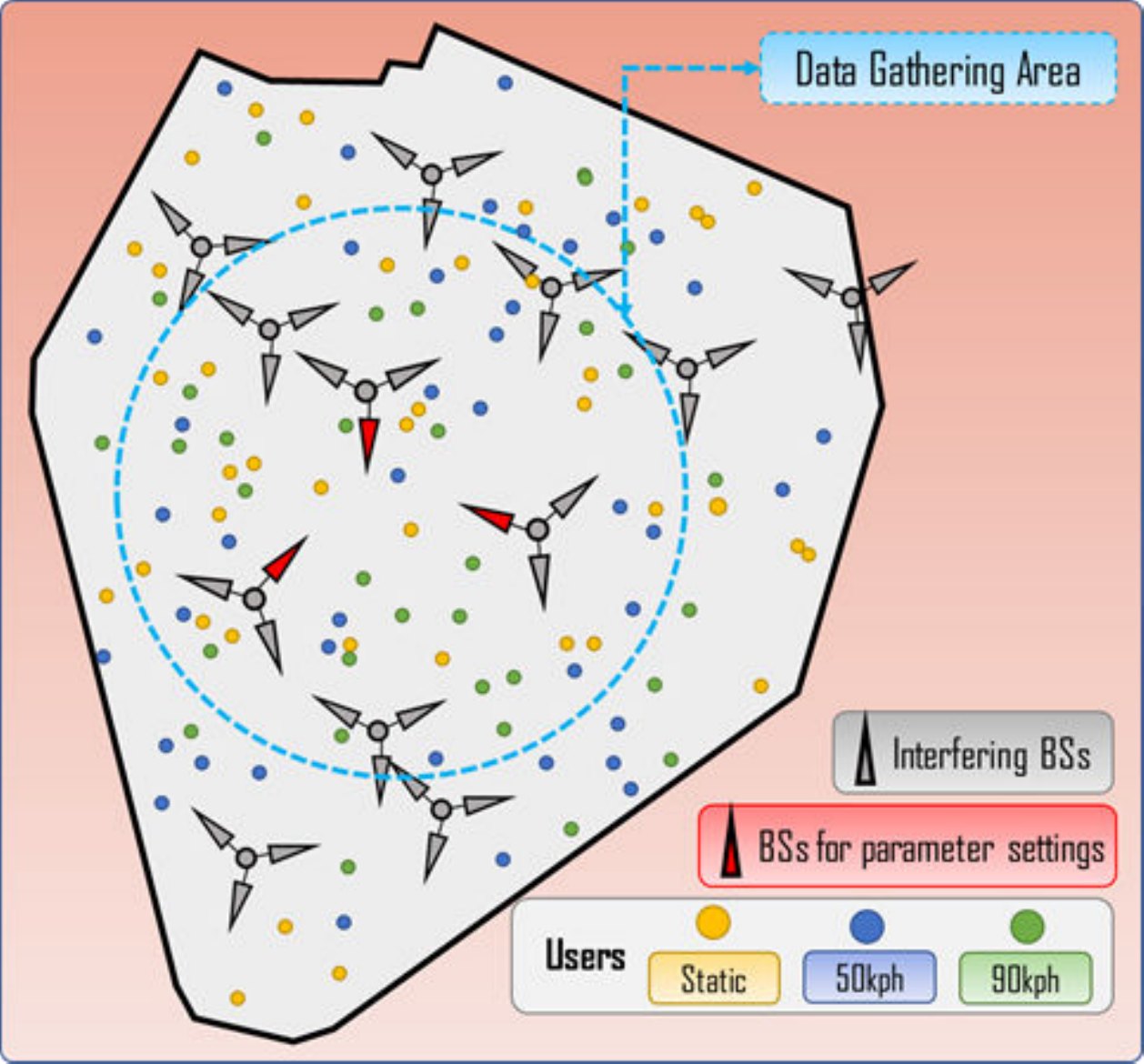}}
\caption{Simulation Environment}
\label{fig}
\end{figure}

In the simulator we defined the serving cell selection process with respect to 3GPP standard specification in terms of qualification, pre-selection and final selection.

\subsubsection{Qualification}\label{AA}
For cells to first be qualified as potential serving cells the RSRP (Reference Signal Received Power) received from the UE must be greater than or equal to the cells' minimum RSRP plus a threshold.
\begin{equation}\ R_{s}^{Tx} \geq T_{s}^{Tx} + Max(0,T_{threshold}^{Tx})\label{eq}
\end{equation}
where $R_{s}^{Tx}$ and $T_{s}^{Tx}$  are the RSRP from the UE and cell's minimum RSRP respectively while $T_{threshold}^{Tx}$ is the cell selection threshold.

\subsubsection{Pre-Selection}\label{AA}
Considering all the potential cells that qualify with the requirements above, only the cell that has the highest RSRP received by the UE is preselected as the serving cell $(S_0)$.

\subsubsection{Final Selection}\label{AA}
Aside from the pre-selected serving cell, any cell among the qualified cells which has the following condition is considered as the best serving cell, that is the highest RSRP plus Cell Individual Offset:
\begin{equation}\ R^{Tx_(c)}  + O_{individual}^{Tx_(c)} \geq R^{S_0} + O_{individual}^{S_0} +M_{HO}^{S_0} \label{eq}
\end{equation}
Here $R^{Tx_(c)}$ and  $O_{individual}^{Tx_(c)}$ are the received power and cell individual offset of the candidate or target cell respectively, $O_{individual}^{S_0}$ is the cell individual offset of the serving cell, $R^{S_0}$ is the received power of the serving cell and $M_{HO}^{S_0}$ is the handover margin. The pre-selected cell is considered the best cell if candidate cell does not fulfill this criterion. 

During the course of the simulation, the UEs will use the cell association process described above in deciding which cell to camp on. Each time the values of \ac{CIO} and \ac{HOM} changes for the 3 BS, the \ac{SINR} from all the users camp on these BS as well as the users within the data gathering area are collected and averaged.

\vspace{-0.3cm}
\begin{table}[htbp]
\caption{PARAMETER SETTINGS FOR SIMULATION}
\begin{center}
\begin{tabular}{|c|c|}
\hline
\textbf{System Parameters }& \textbf{Value} \\
%\cline
\hline
No. of Macro Base Stations & 12 \\
\hline
No. of sectors per Base station & 3\\
\hline
No. of BS for parameter settings & 3 \\
\hline
Carrier Frequency& 2100MHz \\
\hline
Transmission Power & 43dBm \\
\hline
Minimum RSRP& -140dBm\\
\hline
Antenna Gain & 18.5dBi\\
\hline
No. of users in simulation region& 356 \\
\hline
Path Loss model & Ray Tracing \\
\hline
CIO range & Max:10dB, Min:-10dB\\ 
\hline
Handover Margin range & Max: 10dB, Min 0dB \\
\hline
\end{tabular}
\label{tab1}
\end{center}
\end{table}
\vspace{-0.3cm}

\section{Results and Analysis}\label{AA}
Machine learning prediction performances are evaluated in terms of Root Mean Square Error (RMSE) as the validation function. Results shown in Fig. 5 indicate that RMSE ranges from 1.1439 dB to 1.4742 dB which shows a good response to the \ac{ML} models. As expected, Linear Regression performed the least because its limitations lies in matching data that is not linear and predicting data that is not within the range of training sets. CatBoost performs the best because it reduces any form of overfitting and categorizes each attribute accurately. However, it is worth noting that DNN will outperform its counterparts if the entire 12 million plus data set was used. Another interesting thing to note is the good performance of the \ac{ML} models even with 10 percent of training data. The average RMSE of all model is 1.2994 dB showing that even with limited data, \ac{ML} predictions are still able to perform with good level of accuracy.

\begin{figure}[htbp]
\vspace{-0.3cm}
\centerline{\includegraphics[scale=0.5]{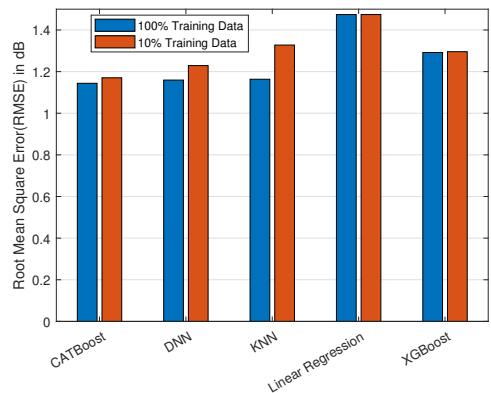}}
\caption{Evaluation of Different \ac{ML} Techniques on \ac{SINR} Prediction.}
\label{fig}
% \vspace{-0.3cm}
\end{figure}

In finding the optimal value of CIO-HOM combination that maximizes SINR, we have evaluated two search techniques, brute force and genetic algorithm. As shown in Fig. 6, both techniques gave close values of maximum SINR. However, it takes brute force 287,000 iterations, which is the same number of samples, to get to an optimum solution. This shows that for large data size, brute force method is computationally inefficient considering the time it takes to search through the entire data set. Meanwhile, \ac{GA} takes considerably less amount of iterations, about 500, to find the optimal solution. This fast convergence of \ac{GA} shows how it can be used efficiently on these kind of problems. \ac{GA} shows that for the three BSs, \ac{CIO} values of [-10, -8, 4] dB combined with \ac{HOM} values of [10, 5, 9] dB will yield the maximum SINR.

Results in Fig. 7 show the efficiency of \ac{GA}. Based from the results, initial configuration of [0, 0, 0] dB for \ac{CIO} and [0, 0, 0] dB for \ac{HOM} for the three BSs give \ac{SINR} of 7.96 dB. However, after 1 generation, which is equivalent to 100 iterations, \ac{SINR} of 9.95 dB is achieved while after 300 iterations, it almost found the optimal combination. This shows that even with less amount of iterations, \ac{GA} can find \ac{SINR} values much faster than brute force approach. This efficiency depends on the generation and population size which involves another optimization operation that is out of scope of this work.

\begin{figure}[t]
\centerline{\includegraphics[width=\linewidth]{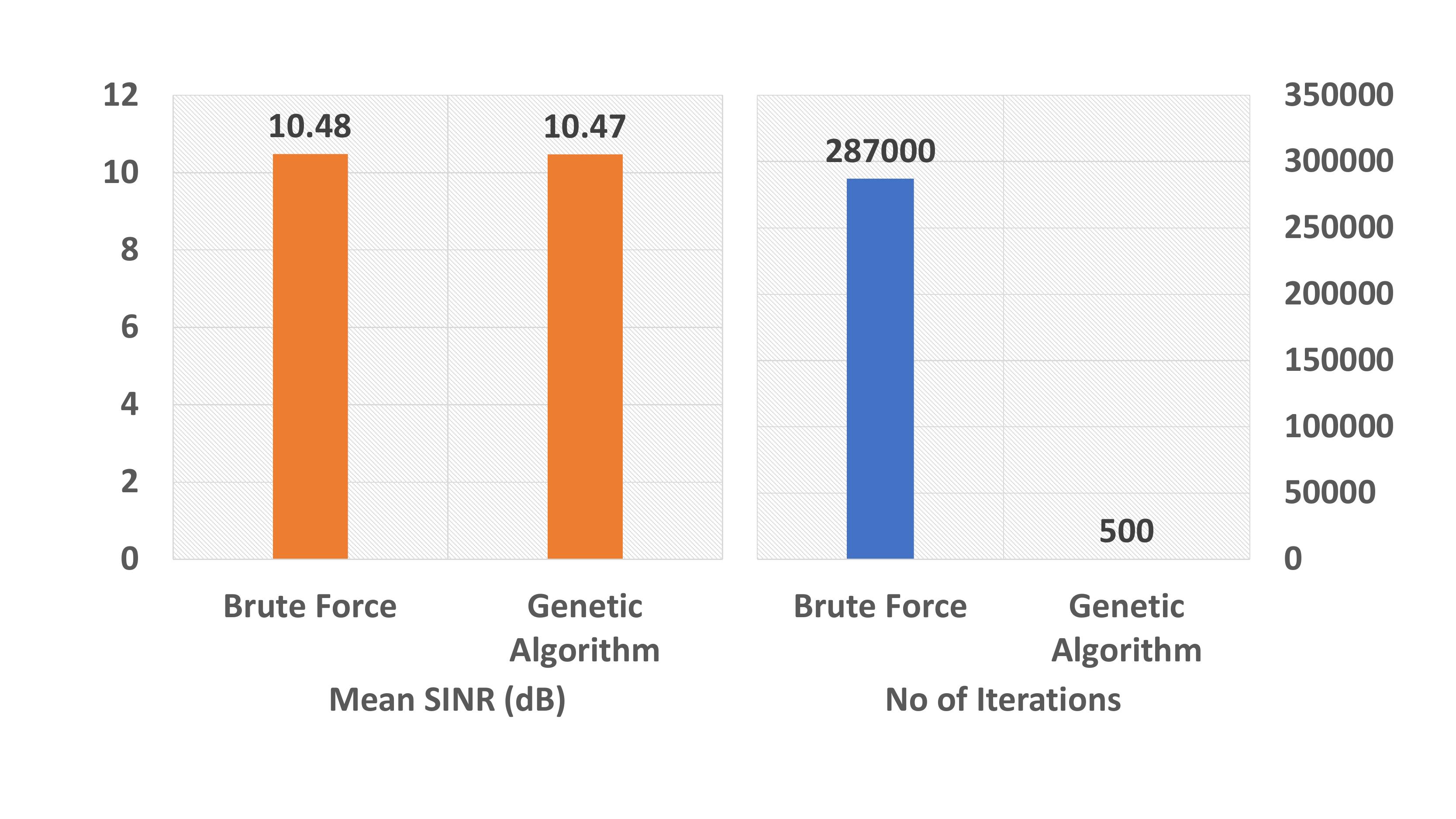}}
\vspace{-0.7cm}
\caption{Optimization Techniques and Evaluation }
\label{fig}
\vspace{-0.3cm}
\end{figure}

\begin{figure}[htbp]
\centerline{\includegraphics[width=0.9\linewidth]{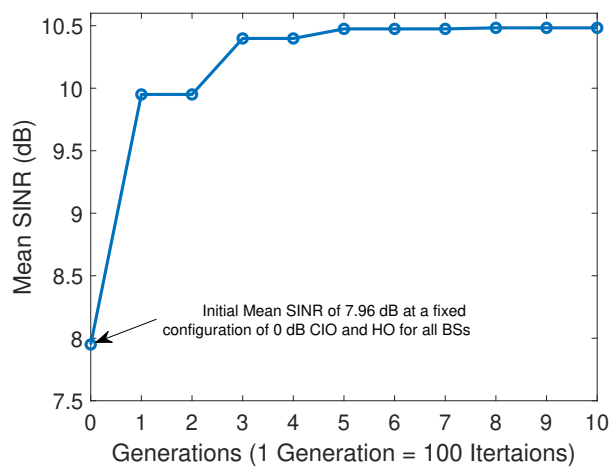}}
\caption{Maximum mean \ac{SINR} vs. Iteration for Genetic Algorithm}
\label{fig}
\vspace{-0.3cm}
\end{figure}
\section*{Conclusion}
In this paper, we present a machine learning based framework to capture \ac{KPI} behavior with changes in COPs and to find the optimal combination of these COPs to maximize the KPI. Using several machine learning techniques, we have predicted the mean \ac{SINR} of all users with different combinations of \ac{CIO} and HOM. Analysis shows that CatBoost performs the best among all other techniques evaluated. Meanwhile, \ac{GA} demonstrates efficiency in finding the optimal CIO-HOM combination that yields maximum mean \ac{SINR}.

For future studies, we will consider the optimization of more COPs with respect to multiple KPIs in a larger network scale involving more base stations.
\section*{Acknowledgment}

This material is based upon work supported by the National Science Foundation under Grant Numbers 1619346, 1559483, 1718956 and 1730650. For more enquires about these projects please visit www.AI4networks.com

\bibliographystyle{ieeetr}
\bibliography{reference}

\end{document}